# A Conceptual Framework for Accountability in Cloud Computing Service Provision

Research in Progress

**Zahir Al-Rashdi**
School of Business Information Technology and Logistics
RMIT University
Melbourne, Australia
Email: zahir.al-rashdi@rmit.edu.au

**Dr Martin Dick**
School of Business Information Technology and Logistics
RMIT University
Melbourne, Australia
Email: martin.dick@rmit.edu.au

**Dr Ian Storey**
School of Business Information Technology and Logistics
RMIT University
Melbourne, Australia
Email: ian.storey@rmit.edu.au

## Abstract

This paper uses a comprehensive review of the academic and professional literature in relation to accountability in the area of cloud computing service provision. It identifies four key conceptual factors that are necessary for an organisation to be considered as accountable. The four factors were found to be: responsibility, assurance, transparency and remediation. A key finding of the paper is that in order to be considered as an accountable cloud service provider, all four factors need to be implemented and be demonstrable by the organisation.

## 1. Introduction

The migration to the cloud has become a global phenomenon (Hobfeld, Schatz, Varela & Timmerer 2012) that has spread widely among government bodies and the private sector. Both sectors share a similar direction and vision for migrating to the cloud, and the issue has received growing attention by both academic and business communities (Zissis & Lekkas 2012). In addition, there is significant evidence that it achieves its goals of flexibility, cost-effectiveness and a proven delivery platform for providing business or consumer IT services over the internet(Pearson 2013). However, there is also a significant amount of agreement about the existence of information security issues prior, during and after cloud implementation, and thus concerns have been voiced about the security issues introduced through the adoption of a cloud computing model (Hashizume, Rosado, Fernández-Medina & Fernandez 2013).

Many researchers have indicated that accountability should be given more attention and treated as a high-priority issue in terms of security (Ko, Jagadpramana, Mowbray, Pearson, Kirchberg, Liang & Lee 2011; Pearson, Tountopoulos, Catteddu, Südholt, Molva, Reich, Fischer-Hübner, Millard, Lotz & Jaatun 2012; Pearson & Wainwright 2013; Rajani, Nagasindhu & Saikrishna 2013; Yao, Chen, Wang, Levy & Zic 2010), as it directly affects the quality of service (QoS) and the grade of service (GoS) (Lee, Tang, Chen & Chu 2012; Ye, Jain, Xia, Joshi, Yen, Bastani, Cureton & Bowler 2010). Most users are seeking assurance (Firdhous, Ghazali & Hassan 2012; Huang & Nicol 2013) that their QoS and GoS requirements are satisfied, and that their operations will not be hindered due to congested cloud resources. Providing the required assurances, measures and guarantees for both QoS and GoS are challenging tasks, and accountability and trust are two major concepts that need to be considered as foundational to potential users embracing cloud services (Chakraborty & Roy 2012; Ferrari 2013; Mouratidis, Islam, Kalloniatis & Gritzalis 2013). Even though technical aspects relating to cloud security and privacy have been actively researched, these conceptual issues have not been addressed in depth.

This paper presents a conceptual framework for understanding accountability in the area of cloud computing service provision. Accountability is a core concern for information security in cloud computing, representing most importantly the trust in service relationships between clients and cloud service providers (CSPs) (Pearson & Wainwright 2013). Without evidence of accountability, a lack of trust and confidence in cloud computing often develops among business management (Ko et al. 2011; Muppala, Shukla & Patil 2012; Pearson 2013; Rashidi & Movahhedinia 2012). It is then considered an added level of risk (Cayirci 2013; Gellman 2012; Guitart, Macias, Djemame, Kirkham, Jiang & Armstrong 2013; Morin, Aubert & Gateau 2012; Rajani, Nagasindhu & Saikrishna 2013), since a client's essential services will be controlled and managed by a third party. Consequently, this new method of outsourcing renders the process of maintaining data security and privacy, supporting data and service availability, and demonstrating compliance far less transparent (Rajani, Nagasindhu & Saikrishna 2013). This makes it difficult for users to understand, influence and determine whether security obligations are actually implemented by CSPs.

A conceptual understanding of the factors relating to accountability in cloud computing service provision is especially important due to significant increases in its usage. According to Columbus (2015), cloud computing adoption in enterprises is rapidly accelerating on a global scale, and it is predicted that by 2018, 59% of the total cloud workloads will be Software-as-a-Service (SaaS) workloads – up from 41% in 2013. In addition, by 2016 over 80% of enterprises globally are likely to be using IaaS, with investments in private cloud computing showing the greatest growth (Columbus 2015). Thus, the significant growth in businesses moving to cloud computing due to the previously mentioned characteristics, in the absence of a specific cloud computing accountability framework, highlights the escalating need for research in this area.

In summary, the main motivation of this paper is to provide an understanding of the conceptual factors that comprise accountability when it is used in the context of cloud computing service provision (Firdhous, Ghazali & Hassan 2012).

## 2.  Background

There are two key concepts which need to be examined in the context of this paper: cloud computing service provision and accountability.

### Cloud computing service provision

According to the National Institute of Standards and Technology (NIST), cloud computing is "a model for enabling ubiquitous, convenient, on-demand network access to a shared pool of configurable computing resources (e.g. networks, servers, storage, applications and services), that can be rapidly provisioned and released with minimal management effort or service provider interaction" (Mell & Grance 2011, p. 2).

Cloud computing is considered one of the most promising technologies in computing today (Popovic & Hocenski 2010). It is dependent on virtualisation technology to simulate physical computers via virtual machines and various physical devices, at a hardware level is used such as processors, hard drives and network devices, are located in data centres (Marston, Li, Bandyopadhyay, Zhang & Ghalsasi 2011). Location independence, resources pooling and rapid elasticity all are critical elements of cloud implementation which depend on virtualisation (Avula, Nela, Gudapati & Velagapudi 2012). In line with this, cloud computing has already addressed a number of key characteristics, such as flexibility/elasticity, scalability of infrastructure, broad network access, location independence, reliability, economics of scale and cost-effectiveness, and sustainability (Zissis & Lekkas 2012).

A cloud model can be implemented in variety of ways depending on the business needs of an enterprise, such as public, private, hybrid or community cloud (Muppala, Shukla & Patil 2012). In addition, clouds can be delivered via three main levels of service: SaaS, Platform-as-a-Service (PaaS) and IaaS (Nayak & Yassir 2012).

Overall, regardless of the significant benefits and rapid growth in its adoption and implementation by a wide range of users, there are three key strategic, operational and people challenges cloud computing still has to confront (Crompton 2015). These challenges are: (1) governance (Nagpal, Patil, Ramanathan & Trevathan 2014); (2) cloud computing environments (Khajeh- Hosseini, Greenwood, Smith & Sommerville 2012) in relation to resource allocation and utilisation based on needs (Xiao, Zhen, Song & Chen 2013), what cloud solution suits an organisation such as a private (Armbrust, Fox, Griffith, Joseph, Katz, Konwinski, Lee, Patterson, Rabkin & Stoica 2010), public (Hofmann & Woods 2010) or hybrid (Mazhelis & Tyrväinen 2012), and which vendors would be right to play each role; and (3) security and privacy (Gartner 2015), meaning how enterprises will ensure accountability in the clouds – that their owner obligations will be accurately implemented in outsourced clouds (Hashizume et al. 2013), and that their data will not be exposed (Wei, Zhu, Cao, Dong, Jia, Chen & Vasilakos 2014).

While there have been many research studies (Wei et al. 2014; Xiao, Zhifeng & Xiao 2013; Yellamma, Narasimham & Sreenivas 2013; Yellamma Pachipala & Challa Narasimham 2014) on cloud computing and information security, the convergence of these two areas encourages further academic efforts towards understanding cloud computing accountability. Some of these topics are discussed as a point along with other issues; thus they are not explored as separate issues. Accountability in cloud service provision is one of the evolving security issues and requires significant attention (Ko et al. 2011).

 In the context of this study, this researcher has taken on the challenge of closely examining accountability in this context, as various scholars (Ko et al. 2011; Ko, Lee & Pearson 2011; Pearson 2011; Pearson & Charlesworth 2009; Pearson et al. 2012; The Centre for Information Policy Leadership 2009a; The Centre for Information Policy Leadership , T 2011; Weitzner, Abelson, Berners-Lee, Feigenbaum, Hendler & Sussman 2008; Yao et al. 2010) believe that this represents the first 'building block' or baseline fundamental that an organisation needs to implement to enhance data protection in the clouds. For example, Ko et al. (2011) stated that accountability in the cloud is one of the evolving issues in cloud security and needs to receive strong attention. In addition, some scholars also believe that accountability is incorporated directly with other information security challenges associated with cloud computing, such as data availability, data privacy and protection, data location and relocation, storage, backup and recovery, loss of physical control, law violation, incompatibility of service, incompatibility of standards, encryption and decryption management, and the cloud computing environment.

Accountability will have a significant impact on the implemented mechanisms (Ko, Lee & Pearson 2011), which in turn will ensure responsible decision-making towards information security management and protection of data (Ko et al. 2011). Thus, this paper will have a great chance to emphasise how central components of accountability are adapted to ensure the accountability approach is correctly implemented for cloud computing.

## What is accountability and how does it relate to cloud computing?

'Accountability' is a common global term that has been extensively used and in computer science, finance and public governance, and is becoming even more incorporated into business regulatory programs. Accountability has also more recently been associated with the concept of a worldwide privacy and data protection framework (Pearson et al. 2012).

The development of the research effort on accountability in relation to cloud computing, has produced a considerable number of definitions, embodying different spheres of accountability research. Both academics and practitioners have different views and interpretations of the accountability concept. For example, accountability in computer science has been referred to as a limited and imprecise requirement that is met by reporting and auditing mechanisms (Cederquist, Conn, Dekker, Etalle & Den Hartog 2005; Pearson 2011); while Yao et al. (2010) considered accountability the way of making the system accountable and trustworthy by the combination of mechanisms. Muppala, Shukla and Patil (2012) referred to accountability as the adherence to accepting the ownership and responsibility towards all actions in a standardized way, as regulated by an acknowledged organisations, such as the Organisation for Economic Cooperation and Development (OECD) which published privacy guidelines in 1980. In addition, Rush (2010) defined accountability as the reporting and auditing mechanisms and obligating an organisation to be answerable for its actions.

In contrast to these definitions, Ko et al. (2011) considered accountability as only one component of trust in cloud computing; the other three are security mechanisms (e.g. encryption), privacy, (the protection of personal or confidential data perhaps not to be exposed), and auditability. Another similar definition for accountability was provided by the Galway project on privacy regulators and privacy professionals, where accountability was defined as the commitment towards safeguarding personal information with an obligation to act as a responsible steward and take responsibility for protecting, managing and appropriate use of that information beyond mere legal requirements, and to be hold accountable for any misuse of that information (The Centre for Information Policy Leadership 2009b).

In addition, the Centre for Information Policy Leadership identified accountability in relation to privacy as "the acceptance of responsibility for personal information protection. An accountable organization must have in place appropriate policies and procedures that promote good practices which, taken as a whole, constitute a privacy management program. The outcome is a demonstrable capacity to comply, at a minimum, with applicable privacy laws. Done properly, it should promote trust and confidence on the part of consumers, and thereby enhance competitive and reputational advantages for organizations" (The Centre for Information Policy Leadership , T 2011, p. 1).

## Evolution of accountability

The principle of accountability is found in a range of regulatory policy' such as the OECD guidelines; in the laws of the European Union (EU), the EU member states, Canada and the United States; and in emerging governance such as the APEC Privacy Framework and the Spanish Data Protection Agency's Joint Proposal for an International Privacy Standard. As an example, the OECD established accountability as a principle of data protection in 1980, and since then it has played an increasingly important and visible role in privacy governance. The emergence of the accountability principle places responsibility on organisations as data controllers to comply with measures that give effect to all of the OECD principles (The Centre for Information Policy Leadership 2009a). In the EU, the principle of accountability initially related to privacy protection, including the implementation of processes by organisations which in turn assess how much data needs to be collected, the usefulness and usability of the collected data for a particular purpose, and the protection level required to ensure information security. The transfer of data outside the EU has been managed with a certain mechanism to ensure the trusted transfer of sensitive and personal data was addressed in the EU accountability principle in data governance section (The Centre for Information Policy Leadership 2010).

'Following on from the EU principle', in February 2009, the Spanish Data Protection Agency established a basis for data transfers and a Joint Proposal for an International Privacy Standard including the principle of accountability (The Centre for Information Policy Leadership, T 2011). At the same time, the office of the Privacy Commission of Canada established the first principle of accountability in 2009 under Canada's Personal Information Protection and Electronic Documents Act (PIPEDA), which soon became part of the law governance mechanism managing the processing, storage and transfer of data domestically and outside Canadian borders (The Centre for Information Policy Leadership 2009b). In the neighbouring United States (US), the government acted initiatively to enhance the principle of data protection and accountability by imposing legal obligations under the Gramm-Leach-Bliley Act. The act's Safeguards Rule, enforced by the Federal Trade Commission, requires financial institutions to have a security plan to protect the confidentiality and integrity of personal consumer information (Weitzner et al. 2008). In addition, the Asia-Pacific Economic Cooperation (APEC) Privacy Framework was established based on the OECD principle to manage the data protection and privacy of personal information intended to be exchanged domestically or internationally, in particular requiring accountability during such data transfers (The Centre for Information Policy Leadership 2009a, 2009b; The Centre for Information Policy Leadership, T 2011).

## 3. Research Methodology

A comprehensive and rigorous review of accountability and its relationship to cloud computing has been conducted focusing on two sources: 1) papers published in both academic and professional literature; and 2) industrial reports published by well-known organisations such as Gartner, Microsoft and The Centre for Information Policy Leadership. This review has concentrated on information security, cloud computing including security issues, information systems, and accountability in cloud computing. When accessing the literature, the researcher used the following keywords to search IEEE Xplore, ScienceDirect, The ACM Digital Library, ProQuest and Google Scholar: cloud computing, cloud computing security, accountability and trust in cloud computing, auditability and trust, confidence, risk assessment and risk management, information security for cloud computing, information security, information systems security, information technology security, information security management, cyber security, information assurance, using SLAs for cloud computing service provision, and information security practices and standards. The preliminary results were sourced from more than 471 scholarly articles, industry standards and technical reports; and a review of abstracts resulted in the elimination of more than 300 that were not related to accountability and cloud computing service provision, leaving the remaining 171.

The main outcomes of this review in relation to accountability for cloud service provision were four common components of accountability which are perceived as the fundamentals that the cloud computing provider needs to demonstrate in order to fulfil its accountability. These four components are transparency, responsibility, assurance and remediation, and have been designed to provide a roadmap for both clients and CSPs. Yet even though these four components are interrelated, it needs to be emphasised that accountability is not a 'one-size-fits-all' approach. The review also showed that each organisation will need to determine which components are more compliant with relevant external criteria such as laws, regulations or industry best practices.

In addition, there are some characteristics that will influence how organisations apply these accountability components, such as the organisation's business model, data holdings, technologies and applications, and the privacy risks they may raise for clients. For example, organisations holding highly sensitive data would implement these accountability components differently to those with less sensitive data. Therefore, every organisation needs to customise how the four accountability fundamentals are implemented in order to demonstrate its accountability. A thorough explanation of these four central components of accountability is included in the following section.

## 4. Conceptual Factors

The Centre for Information Policy Leadership identified accountability as "a demonstrable acknowledgement and assumption of responsibility for having in place appropriate policies and procedures, and promotion of good practices that include correction and remediation for failures and misconduct. It is a concept that has governance and ethical dimensions. It envisages an infrastructure that fosters responsible decision-making, engenders answerability, enhances transparency and considers liability. It encompasses expectations that organisations will report, explain and be answerable for the consequences of decisions about the protection of data. Accountability promotes implementation of practical mechanisms whereby legal requirements and guidance are translated into effective protection of data" (The Centre for Information Policy Leadership 2010, p. 2)

In this context, this paper uses the definition of accountability offered by (Muppala, Shukla & Patil 2012; The Centre for Information Policy Leadership 2009b; The Centre for Information Policy Leadership , T 2011), as identified above. In particular, these definitions cover the central components of accountability: transparency, responsibility, assurance and remediation. Those definitions are used as the basis of this study, because they incorporate the protection of organisational assets and personal privacy.

*Figure 1* shows the overall interactions between these four key components in terms of achieving accountability. It should be noted that the double arrows indicate that the four factors interact with each other and are not necessarily independent of each other.

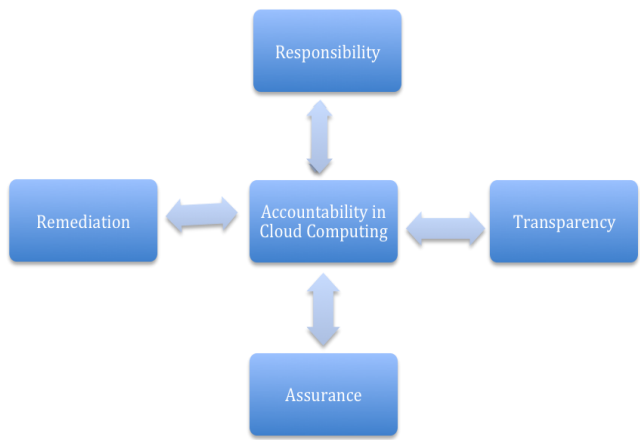

*Figure 1: The four central components of accountability*

## Responsibility

Responsibility in the context of accountability for cloud computing service provision is the acknowledgment and assumption of responsibility by CSPs that they have introduced or have in place appropriate policies and procedures (Ko, Lee & Pearson 2011). Responsibility would be achievable by ensuring the existence of obligatory and enforceable written data privacy policies and procedures that reflect applicable laws, regulations and industry standards (Takabi, Joshi & Ahn 2010). The accountable CSP should develop, implement and communicate to clients a set of data privacy policies that are informed by appropriate external criteria recognised by laws, regulations or industry best practices (Toney & Kadam 2013). In addition, the accountable CSP should be prepared to provide clients with (Wang, Wang, Ren & Lou 2010) and also design and deploy a set of procedures to implement effective and practical written policies according to the circumstances of each organisation, such as what data is collected, how it is used, and how systems and organisations are connected (Weitzner et al. 2008).

Responsibility by CSPs involves executive oversight; that is, internal oversight and responsibility for data privacy and protection (The Centre for Information Policy Leadership , T 2011). This can be achieved via the recruitment of a data privacy leader equipped with suitable resources and personnel, who will be responsible for reporting to the CSP organisation's leadership (Cederquist et al. 2005). Senior management should be committed to supporting this process via: (1) appropriate reporting and oversight of the organisation's privacy program; and (2) recruitment of senior-level executives in order to develop and implement the CSPs programs, policies and practices (Buyya, Yeo, Venugopal, Broberg & Brandic 2009). However, small- and medium-sized organisations should allocate such oversight resources according to their organisational context, such as the extent and sensitivity of their data holdings and the nature of the data usage (Rajani, Nagasindhu & Saikrishna 2013). By fulfilling the responsibility component, CSPs will have a stronger impact on client's side by increasing the trust and confidence towards migrating businesses to cloud services, which in turn will increase the accountability and reduce the risk (Ko et al. 2011).

Responsibility is considered one of the most important factors of accountability to adequately manage the relationship between CSPs and clients. The accountable organisation (CSPs) should have a data privacy program in place to establish, demonstrate and test its accountability. Each organisation should demonstrate a level of responsibility and willingness to be accountable for any misconduct in its data practices, policies and procedures, which should be implemented based on external legislative criteria, generally accepted principles or industry best practices. All policies and procedures must be approved at the highest level of the organisation, and senior management should demonstrate their commitment towards motivating the responsibility, which in turn will encourage accountability.

Globally the EU Directive, OECD guidelines and APEC principles have established policies and procedures in accordance with legislative, in order to demonstrate their commitment to the implemented policies and internal practices.

## Assurance

The assurance approach in terms of accountability for cloud for cloud computing service provision is to comply with governance and ethical measurements along with promoting the implementation of practical mechanisms are commonly considered key parts of accountability processes and procedures (The Centre for Information Policy Leadership 2010). The following are part of assurance in relation to accountability for cloud computing service provision.

### Staffing and delegation

Adequately trained personnel will ensure the validity of the CSPs privacy program, and will appropriately allocate the right resources to the right staff (Nunez, Fernandez-Gago, Pearson & Felici 2013). Each organisation should have a reasonable number of staff to ensure their privacy program runs smoothly, providing adequate training to enable this and to also incorporate any new developments in the organisation's business model, such as data collection practices and technologies, and consumer offerings (Popović 2010). The appropriate delegation of authority and responsibility for data protection within an organisation is deemed both effective and workable to an accountable CSP. Small- and medium-sized organisations should ensure such delegations align with their specific activities and circumstances, such as the nature, size and sensitivity of their data holdings (Li, Yu, Ren & Lou 2010). When correctly implemented, the client and CSP relationship will be more assured; with CSPs conveying their employees as appropriately trained and assigned responsibility for the privacy program (Pearson et al. 2012).

### Education and awareness

An efficient education and awareness program will ensure that an organisation's personnel and on-site contractors are kept up-to-date on data protection obligations (Ramgovind, Eloff & Smith 2010). Such continued education and awareness will enhance CSP employees' capabilities and increase their understanding of the essentiality of protecting clients' data to avoid data leakage consequences (Bristow, Dodds, Northam & Plugge 2010) such as job dismissal (Ko et al. 2011). This process, without any doubt, will increase the level of trust between clients and their CSPs (Ko, Lee & Pearson 2011)

### Ongoing risk assessment and mitigation

To be an accountable CSP, processes are needed to understand the related risks to privacy that could occur from the implementation of new solutions, products, services, technologies or business models (Pearson 2011). The outcomes of ongoing risk assessment and mitigation (Zhang, Wuwong, Li & Zhang 2010) should be reflected in the organisation's measures taken to mitigate clients' potential risks. They should not treat risk assessment as static; rather that it is dynamic and evolving according to the nature of the collected data, and the usage and processing of this data (Jansen & Grance 2011). Organisations should also demonstrate a satisfactory level of risk analysis in relation to their organisational context to be accountable in risk assessment and mitigation practices (Heimgartner 2015). In addition, further demonstration should be made by these organisations of how such decisions are made and the steps taken to mitigate risk (Clemons & Chen 2011). By having such steps in place including precise processes and procedures to arrest ongoing risk and mitigation, CSPs will be positively perceived by clients in terms of accountability, and are likely to be branded as a trusted party (Ko, Kirchberg & Lee 2011)

### Program risk assessment oversight and validation

Constant reviews of an organisation's privacy and data protection accountability program should be considered by both clients and CSPs to be perceived as an accountable organisation (Jansen & Grance 2011). This will ensure the consistent meeting of an organisation's needs via sound decisions on data management and protection that successfully promote and meet the privacy outcomes (Clemons & Chen 2011). An exchange review of such programs between CSPs and clients will improve trust and validate CSP programs, which in turn conveys an accountable partnership (Khajeh- Hosseini et al. 2012)

### Event management and complaint handling

An accountable organisation should have in place procedures that effectively respond to inquiries, complaints and violations of data protection (The Centre for Information Policy Leadership 2010). A timely response to any inquiries, complaints or violations in terms of data protection will establish an image of support between clients and CSPs, which in turn will strengthen the accountability prospective between both of them (Bristow et al. 2010). The support matter between clients and their CSPs can sometimes become complicated, which often leads to taking the contract off from the third party, which in short means a loss of trust between clients and CSPs (Muppala, Shukla & Patil 2012).

### Internal enforcement

Accountable organisations should have in place methods to enforce internal policy, ensuring that any breaches to those internal data protection rules by employees, such as misappropriate or misuse of data, are subject to sanctions including discharge (The Centre for Information Policy Leadership 2010). In this case, such internal enforcement is directly connected to CSPs as the third party or contractors, and strengthening these enforcement aspects will enhance the likelihood of clients choosing to employ that CSP (Bristow et al. 2010).

Overall assurance and accountability are interconnected. Based on the business culture, each organisation is expected to establish performance systems to be perceived as an accountable organisation, and the following characteristics represent successful performance systems: (1) they are consistent with the organisation's culture and are integrated into business processes; (2) they assess risk across the entire data life cycle; (3) they include training, decision-making tools and monitoring; (4) they apply to outside vendors and other third parties, to ensure personal data obligations are met no matter where the data is processed; (5) they allocate resources where the risk to individuals is greatest; and (6) they are a function of an organisation's policies and commitment. In Europe, North America and Asia-Pacific seal programs are used where they are the online third party accountability agents, which provide external oversight by making assurance and verification reviews a requirement for participating organisations.

## Transparency

In order to ensure transparency in the context of accountability for cloud computing service provision a range of issues need to be covered. An outcome of each review including modifications to rules and procedures should be made available to clients in a clear and timely manner (Takabi, Joshi & Ahn 2010). The information should be appropriately conveyed to both client organisations and regulators in a rigorous and cost-effective manner (Pearson & Charlesworth 2009). As part of this process, the outcome of assessment measures or audits should be reported to the appropriate employee within an client organisation, and if necessary corrective action should be taken (Patel, Ranabahu & Sheth 2009). Transparency includes reports and explanations of decisions that have been made to protect data. It also means that acceptance of liability and redress actions are clearly presented to clients (Muppala, Shukla & Patil 2012). The transparency between clients and their CSPs is an essential element towards achieving accountability in cloud computing, as most clients want to know who is handling the data, how it is used, and where and when. To exchange such information between clients and CSPs will increase confidence and trust, which in turn will increase accountability (Ko et al. 2011).

A level of transparency should be demonstrated between clients and organisations. Clients should have the right over the collected data by checking them, and stop using certain data where they are inappropriate or alternatively to correct the collected data when they are inaccurate. However, there may be limitations to the disclosure of information in some circumstances.

## Remediation

To achieve remediation that ensures accountability for cloud computing service provision, there is a need for a range of remediation processes. Remediation according to The Centre for Information Policy Leadership (2010, p. 7) is "the method by which an organisation provides remedies for those whose privacy has been put at risk". In this context, an accountable organisation should promote a best practice in correction and redress in the case of failure and misconduct (Pearson 2011). In addition, an accountable organisation should have a specific remediation mechanism that suits each organisation according to their data holdings, and the way the data is used and appropriate for a specific issue. These mechanisms should be readily and easily accessible to clients, and be able to address complaints in an effective and efficient manner (Pearson 2013). The redress mechanisms would be different from culture to culture and from industry to industry; decisions about redress would be made locally.

However, these remediation mechanisms would need to be developed in consultation with a range of experts, regulators, civil society, and representatives of both public and private sector organisations (Jansen & Grance 2011). To establish a credible accountability approach, CSPs' redress mechanisms should be fully discussed and examined, with each mechanism adequately serving the requirements of all related parties such as national culture, regulations, self-regulations and laws (Pearson 2011). Such mechanisms would increase the confidence between clients and CSPs, and strengthen the decision on migrating a business to a cloud environment, which in turn will satisfy the accountably approach (Takabi, Joshi & Ahn 2010).

Remediation is complementing the accountability process and procedures where an organisation's and clients to ensure business continuity in case of failure should undertake a set of tasks. When failure occurs, individuals should have access to a recourse mechanism. For instance a third party agency might be needed to address and resolve the failure that has occurred. Clients should be aware of the process and procedures to be followed in case of failure.

## Summary

As detailed above, the literature revealed four main components of accountability in relation to cloud computing service provision, and shows that in order to be an accountable organisation they should be fully implemented. As can be seen in the descriptions of these factors, there are inherent interactions between the four components (responsibility, assurance, transparency and remediation) which mean that they should all be addressed simultaneously; individually implementing each component will likely cause failure in relation to accountability. Cloud service providers that wish to be regarded as accountable should be aware of these four fundamental accountability components, and must be prepared to demonstrate to clients their accomplishment of them. However, this does not mean that there is a singular method for implementing all four factors or any individual factor – the nature of the organisation, its industry context, the type of data collected, the business model, and the potential risks that data usage raises for clients all have an impact on the method chosen for implementing these factors (Pearson 2011).

## 5. Conclusion

This paper has sought to understand how accountability in cloud computing can be conceptualised. The wide range of existing research into accountability in cloud computing has used a technical approach and has been quantitative, and has generally not addressed the conceptual issues. The enormous growth in moving businesses to cloud computing, mainly due to its flexibility, cost-effectiveness and scalability, and the corresponding absence of a specific cloud computing accountability framework, highlights the growing need for research in this area. This study has used an extensive analysis of the literature relating to cloud computing and accountability for information security to develop a model of the key conceptual factors (transparency, responsibility, assurance and remediation) relating to this issue.

## 6. Future Work

This paper is part of an ongoing research program, and it is planned to conduct a series of case studies to examine the real-life experiences of organisations in ensuring accountability when they adopt cloud computing service provision, using those experiences to analyse the four conceptual factors that underlie accountability. The findings of this research are expected to contribute to the growing awareness of the importance of accountability, in order to understand the importance of the accountability factors that motivate policymakers to adopt cloud computing projects and migrate their business to the cloud prior to, during and after their implementation, and to evaluate what happened and why in terms of security issues. This study is aiming to produce a holistic and heuristic accountability framework that enables a theoretical-based description and analysis of the gap that exists between the ideal and the actuality of the institutional and technical environments of cloud computing implementation in regards to accountability. Finally, this study is seeking to find ways to strengthen the trust and confidence between organisations that have adopted the cloud and CSPs, which in turn strengthen the accountability, which also helps to reduce risks connected to the adoption of cloud computing services.

# Copyright